\documentclass[12pt,preprint]{aastex}

\usepackage{amsmath}

\slugcomment{published 1 May 2013, {\it ApJ}, {\bf 768}, 43}

\shorttitle{Concentric Maclaurin Spheroids}
\shortauthors{Hubbard}

\begin{document}

\title{Concentric Maclaurin spheroid models of rotating liquid planets}

\author{W. B. Hubbard}
\affil{Lunar and Planetary Laboratory, University of Arizona,
    Tucson, AZ 85721}

\begin{abstract}
I present exact expressions for the interior gravitational potential $V$
of a system of $N$ concentric constant-density (Maclaurin) spheroids.
I demonstrate an iteration procedure
to find a self-consistent solution for the shapes of the
interfaces between spheroids, and for the interior
gravitational potential.  The external free-space potential,
expressed as a multipole expansion, emerges
as part of the self-consistent solution.  The procedure is both simpler and
more precise than perturbation methods.  One can choose the distribution
and mass densities of the concentric spheroids so as to
reproduce a prescribed barotrope to a specified accuracy.
I demonstrate the method's efficacy by comparing its results
with several published test cases.
\end{abstract}


\section{Introduction}

In its general form, the problem of the theory of figures is to find the external
gravitational potential of a liquid planet in hydrostatic equilibrium, rotating at
a uniform rate $\omega$, and obeying a specified
barotropic relationship for the dependence of pressure $P$
on mass density $\rho$.  

The expected precision ($\sim$ one part in $10^9$)
of the $\it {Juno}$ Jupiter orbiter spacecraft's measurements of Jupiter's gravity
field will require a gravitational-modeling theory of unprecedented accuracy
\citep{kas10}.
\citet{WBH12}   (Paper I) is an initial step
toward such a theory.  Paper I presents a new approach to
the calculation of the multipole expansion of the external gravitational potential of
a rotating planet in hydrostatic equilibrium.

As is well known, the problem of the theory of figures
can be solved in closed
or partially-closed form for a small number of special barotropes
but arbitrary barotropes generally require numerical methods.
Analytic methods balloon in complexity even
for the apparently simple case of two constant-density
layers, the so-called two-layer Maclaurin spheroid \citep{schu11, kong12}.  In
principle, such analytic complexity could be bypassed
by seeking a purely numerical
solution to the general equation of hydrostatic equilibrium.  However,
numerical solutions are vulnerable to numerical noise,
produced for example by cancellation of nearly equal terms.
In the traditional approach,
cancellation of terms is mitigated by solving a hierarchy
of integrodifferential equations generated from a perturbation
expansion of the mass distribution in powers of the rotation rate
\citep{zt78}.

Paper I shows how the particular problem of
the hydrostatic equilibrium of a rotating
constant-density planet can be numerically solved to high precision by
using gaussian quadrature to obtain the mass multipole moments.
In this method, the moments are calculated by
performing one-dimensional integrals over the surface mass
distribution.  Although gaussian quadrature is a numerical approximation
to analytic integration, the results are exact (to within the floating point precision
of the computer), as long as the integrand can be
expressed as a polynomial of degree
less than the degree of the gaussian quadrature; for a Maclaurin
spheroid, using this approach with 48 quadrature points yields
numerical results with a precision of at least $\sim 10^{-12}$.
The mass multipole moments are then
self-consistently iterated on the shape of the surface (Paper I).  The method
of Paper I largely bypasses the analytic complexity of perturbation methods and
avoids cancellation problems, but as presented is only valid for
a constant-density object, or for a constant-density object with special boundary
conditions.

The present paper shows how the method of Paper I is straightforwardly
generalized to solve the problem of multiple-layered constant-density spheroids.
The resulting method, which I call the concentric Maclaurin spheroid (CMS) method,
retains all of the advantages of the approach of Paper I, with the additional
flexibility that the concentric Maclaurin spheroids can be arranged in sufficient
numbers to closely approximate any prescribed barotrope.  As we will see, an
actual density discontinuity such as a discrete core or first-order phase transition
is trivially incorporated in the CMS method,
as opposed to the usual theories of figures.

In the following Section 2, I present the analytic development of the CMS method.
In Section 3, I apply the CMS method to several published test cases and I show
how the method can incorporate a prescribed barotrope.  In the conclusion (Section 4),
I discuss how the CMS method can be applied to analysis of {\it Juno}
gravity data expected to arrive beginning in 2016.

\section{Theory for $N$ Layers of Maclaurin Spheroids}

\subsection{Exact calculation of gravitational potential}

Consider a configuration of $N$ concentric Maclaurin spheroids (Fig. 1).  Label
the spheroids with index $i=0, 1, \mathellipsis , N-1$, with $i=0$ corresponding to the
outermost spheroid and $i=N-1$ corresponding to the innermost.

\begin{figure}
\epsscale{.80}
\plotone{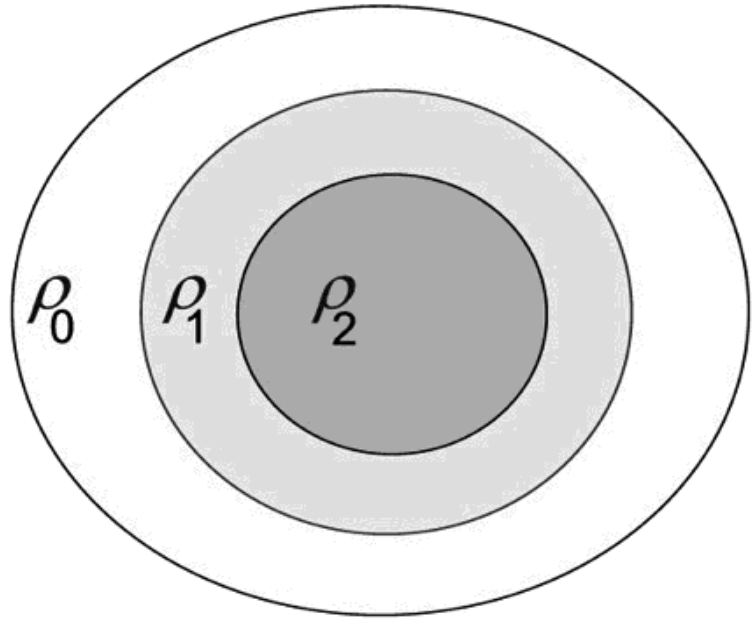}
\caption{Concentric Maclaurin spheroids, each layer with constant density $\rho$, for the case $N=3$.}
\label{fig1}
\end{figure}

Because the gravitational potential $V$ is linear in
the mass density $\rho$, we may use
the principle of superposition, such that the total potential at any point in space
is the sum of the partial potentials of $N$ concentric constant-density spheroids.
Figure 2 illustrates this concept for a three-layer model.

\begin{figure}
\epsscale{.80}
\plotone{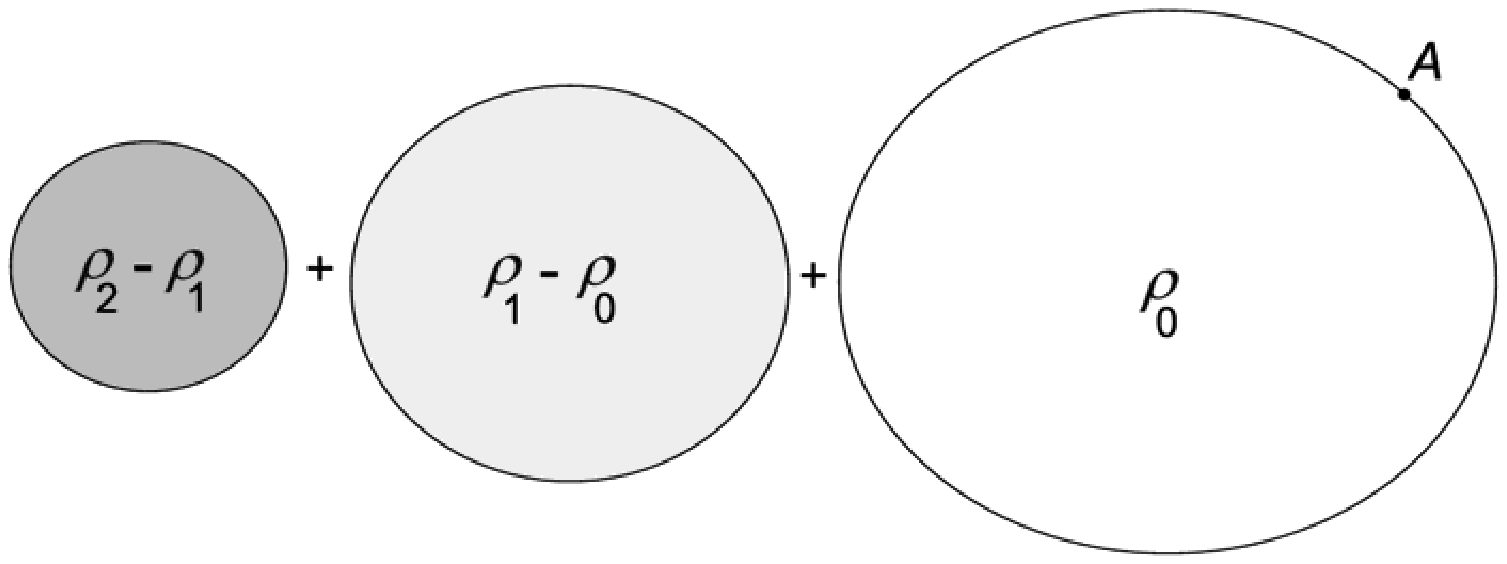}
\caption{Method of superposition of Maclaurin spheroids, for the case $N=3$. 
 The point ``A'' is a typical point on the outermost surface of the planet.}
\label{fig2}
\end{figure}

Let the equatorial radius of the outermost spheroid be $a_0$,
and let the equatorial radii of
the concentric spheroids be $a_0 > a_1 > a_2 > \mathellipsis > a_{N-1}$.  The total external gravitational
potential at some point "A" on the outermost level surface is
\begin{align}
V_{{\rm ext}, A} &= {G \over r}  \left(   \sum_{k=0}^{\infty} D_{0,2k} r^{-2k} P_{2k}(\mu)
  + \sum_{k=0}^{\infty} D_{1,2k} r^{-2k} P_{2k}(\mu) +\mathellipsis \right. \nonumber\\
  &+ \left. \sum_{k=0}^{\infty} D_{N-1,2k} r^{-2k} P_{2k}(\mu) \right),
\end{align}
where $r$ is the radius from the center of the planet, $\mu$ is the cosine of the angle from the rotation axis, the $P_{2k}(\mu)$ are the usual Legendre polynomials,

\begin{equation}
D_{0,2k} = {{2 \pi \rho_0} \over {2k+3}} \int_{-1}^{1} d \mu \, P_{2k}(\mu) \, r_0(\mu)^{2k+3},
\end{equation}

\begin{equation} \label{d12k}
D_{1,2k} = {{2 \pi (\rho_1 - \rho_0)} \over {2k+3}} \int_{-1}^{1} d \mu  \, P_{2k}(\mu) \, r_1(\mu)^{2k+3},
\end{equation}
etc., where the relation $r_i = r_i(\mu)$ is the surface equipotential of the $i$-th layer.  The zero-degree values are
given by

\begin{equation}
D_{0,0} = {{2 \pi \rho_0} \over {3}} \int_{-1}^{1} d \mu \, r_0(\mu)^{3},
\end{equation}

\begin{equation}
D_{i,0} = {{2 \pi (\rho_i - \rho_{i-1})} \over {3}} \int_{-1}^{1} d \mu \, r_i(\mu)^{3},
\end{equation}
and so we have for the total mass $M$

\begin{equation}
M = \sum_{i=0}^{N-1} D_{i, 0}.
\end{equation}

We now introduce the usual dimensionless forms of the multipole moments,

\begin{equation}
M a_0^{2k} J_{i,2k} = -D_{i, 2k}.
\end{equation}
and the dimensionless radii of level surfaces
\begin{equation}
\xi_i = r_i(\mu) / a_0.
\end{equation}
The total external gravitational potential at point ``A'' can thus be rewritten
\begin{equation} \label{Vext_def2}
V_{{\rm ext}, A} = {GM \over r} \left(1 -  \sum_{i=0}^{N-1} \sum_{k=1}^{\infty} J_{i,2k} \,\xi_0(\mu)^{-2k} P_{2k}(\mu) \right),
\end{equation}
where
\begin{equation} 
J_{i,2k} = -\left({3 \over {2k+3}} \right)
\left( {{\delta \rho_i \int_0^1 d\mu \, P_{2k}(\mu) \, \xi_i (\mu)^{2k+3} }
\over { \sum_{j=0}^{N-1}  \delta \rho_j \int_0^1 d\mu \, \xi_j (\mu)^{3} } }\right)  ,
\end{equation}

\begin{equation} 
\delta \rho_i = \rho_i - \rho_{i-1}
\end{equation}
for $i > 0$ and
\begin{equation} 
\delta \rho_0 = \rho_0.
\end{equation}

Next, we must compute the total gravitational potential on
an interior interface (level surface) at an arbitrary point ``B'', as
shown in Fig. 3.

\begin{figure}
\epsscale{.80}
\plotone{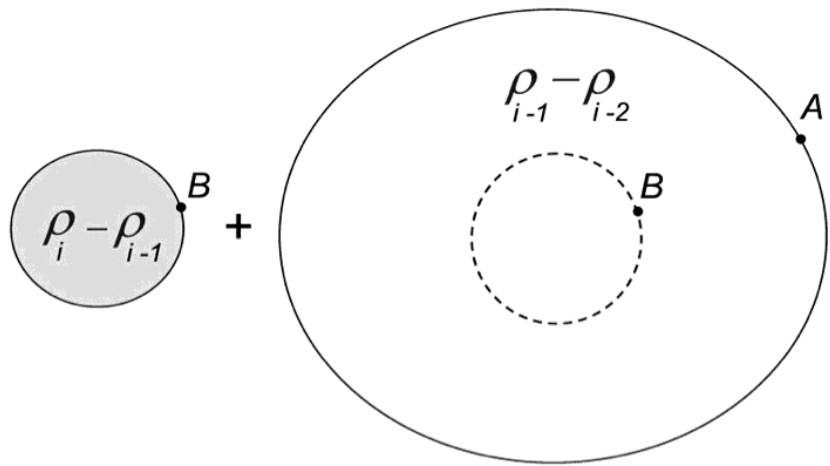}
\caption{Schematic diagram illustrating the computation of three
contributions to the gravitational potential at point ``B'' on an interior interface.}
\label{fig3}
\end{figure}

First, we consider a problem in which there is only a mass distribution
with a constant density $\delta \rho_i$ interior to point ``B'' located at
coordinates $(r,\mu)$.  We calculate the external potential due to this mass distribution, finding
\begin{equation}
V_{i,{\rm ext}, B} = {G \over r}  \sum_{k=0}^{\infty} D_{i,2k} r^{-2k} P_{2k}(\mu) .
\end{equation}
Next we calculate the external potential at the surface of a spherical mass
distribution with radius $r$ and constant density $\delta \rho_{i-1}$
(shown as a dashed circle in Fig. 3):
\begin{equation}
V''_{i,{\rm ext}, B} = {G \over r}  { {4 \pi \delta \rho_{i-1}} \over 3} r^3.
\end{equation}
Finally, we calculate the internal potential at point ``B'' due to the mass distribution
with constant density $\delta \rho_{i-1}$ external to the dashed circle in Fig. 3:
\begin{equation} 
V'_{i,{\rm int}, B} = {2 \pi G \over r} \delta\rho_{i-1}  \sum_{k=0}^{\infty}  P_{2k}(\mu) 
\int_{-1}^1 d \mu'  \, P_{2k}(\mu') \int_r^{r_{i-1}(\mu')} dr' \, {r'}^{-2k+1} .
\end{equation}

Adding all contributions to the potential at point ``B'' due to the mass density in
the $i$-th layer and in the $i-1$-th layer, one has
\begin{equation} 
V_{i, B} = {G \over r}  \sum_{k=0}^{\infty} D_{i,2k} r^{-2k} P_{2k}(\mu)
  + G \sum_{k=0}^{\infty} D'_{i-1,2k} r^{2k} P_{2k}(\mu) +
GD''_{i-1,0} r^2,
\end{equation}
where [{\it cf} Eq. (4)]
\begin{equation} 
D_{i,2k} = {{4 \pi \delta\rho_i} \over {2k+3}} \int_{0}^{1} d \mu \, P_{2k}(\mu) \, r_i(\mu)^{2k+3},
\end{equation}
for $k > 1$ we have
\begin{equation} 
D'_{i-1,2k} = {{4 \pi \delta\rho_{i-1}} \over {2-2k}} \int_{0}^{1} d \mu \, P_{2k}(\mu)\, r_{i-1}(\mu)^{2-2k},
\end{equation}
for $k = 1$ we have
\begin{equation} 
D'_{i-1,2} = 4 \pi \delta\rho_{i-1} \int_{0}^{1} d \mu  \, P_{2}(\mu) \ln [r_{i-1}(\mu)],
\end{equation}
and for $k = 0$
\begin{equation} 
D'_{i-1,0} = 2 \pi \delta\rho_{i-1} \int_{0}^{1} d \mu \, r_{i-1}(\mu)^2,
\end{equation}
\begin{equation} 
D''_{i-1,0} = -{{2 \pi \delta\rho_{i-1}} \over 3}.
\end{equation}

Next, analogous to Eq. (7), we introduce dimensionless forms of the $D'_{i-1,2k}$:
\begin{equation}
M a_0^{-1-2k} J'_{i,2k} = -D'_{i, 2k},
\end{equation}
and
\begin{equation}
J''_{i,0} = {{2 \pi \delta \rho_i a_0^3} \over {3M}}.
\end{equation}

By analogy with Eq. (10), we may write the dimensionless forms
of Eqs. (18-21): for $k>1$
\begin{equation} 
J'_{i,2k} = -\left({3 \over {2-2k}} \right)
\left( {{\delta \rho_i \int_0^1 d\mu \, P_{2k}(\mu) \, \xi_i (\mu)^{2-2k} }
\over { \sum_{j=0}^{N-1}  \delta \rho_j \int_0^1 d\mu \, \xi_j (\mu)^{3} } }\right)  ,
\end{equation}
for $ k = 1$
\begin{equation} 
J'_{i,2} = -3
\left( {{\delta \rho_i \int_0^1 d\mu \, P_{2}(\mu) \ln[\xi_i (\mu)] }
\over { \sum_{j=0}^{N-1}  \delta \rho_j \int_0^1 d\mu \, 
\xi_j (\mu)^{3} } }\right)  ,
\end{equation}
and for $k = 0$
\begin{equation} 
J'_{i,0} = -\left({3 \over 2} \right)
\left( {{\delta \rho_i \int_0^1 d\mu \, P_{2k}(\mu) \, \xi_i (\mu)^{2} }
\over { \sum_{j=0}^{N-1}  \delta \rho_j \int_0^1 d\mu \, 
\xi_j (\mu)^{3} } }\right)  ,
\end{equation}
\begin{equation} 
J''_{i,0} = 
 { \delta \rho_i 
\over {2 \sum_{j=0}^{N-1}  \delta \rho_j \int_0^1 d\mu \, \xi_j (\mu)^{3} } }  .
\end{equation}
Eq. (16) then takes the form
\begin{equation} 
V_{i, B} = -{GM \over a_0} {1 \over \xi}  \left[ \sum_{k=0}^{\infty} J_{i,2k} \xi^{-2k} P_{2k}(\mu)
  + \sum_{k=0}^{\infty} J'_{i-1,2k} \xi^{2k+1} P_{2k}(\mu) +
J''_{i-1,0} \xi^3 \right].
\end{equation}

The total potential at a point B
located at coordinates $(\xi,\mu)$ on the $j$-th interface
is obtained by summing over all layers:
\begin{eqnarray} 
V_{B}(j) = -{GM \over a_0} {1 \over \xi} 
 \left[ \sum_{i=j}^{N-1} \sum_{k=0}^{\infty} J_{i,2k} \xi^{-2k} P_{2k}(\mu)
 \right. \nonumber\\
 \left. + \sum_{i=0}^{j-1} \sum_{k=0}^{\infty} J'_{i-1,2k} \xi^{2k+1} P_{2k}(\mu) +
   \sum_{i=0}^{j-1} J''_{i-1,0} \xi^3 \right].
\end{eqnarray}

\subsection{Parameters and scaling}

Assume that the planet rotates as a solid body at an angular
rate $\omega$.  Therefore in the corotating frame there appears a
rotational potential
\begin{equation} 
Q = {1 \over 3} r^2 \omega^2 [1 - P_{2}(\mu)],
\end{equation}
and the total potential $U$ appearing in the equation of hydrostatic equilibrium
\begin{equation} 
\nabla P = \rho \nabla U
\end{equation}
is given by
\begin{equation} 
U = V + Q.
\end{equation}
For a nonrotating planet, all multipole moments for $k > 0$ vanish
and the potential $V$ within the planet depends only on $r$.  The presence of
the nonspherical term $Q$ in $U$ breaks the spherical symmetry and excites
all of the $k > 0$ terms.  We represent the magnitude of $Q$ by the
dimensionless parameter

\begin{equation} 
q = {{\omega^2 a_0^3} \over {GM}}.
\end{equation}

The number and location of the concentric Maclaurin spheroids can be chosen
arbitrarily.  Let the equatorial radius of the $i$-th spheroid be $a_i$.  Let
\begin{equation} 
\lambda_i = {{a_i} \over {a_0}}, i = 0, 1, \mathellipsis, N-1.
\end{equation}
The $\lambda_i$ can be spaced equally between $0$ and $1$, or could be made denser
in certain regions (for example, one could space them at two or three per density
scale height).

Define the mean density of the planet:
\begin{equation} 
\bar{\rho} = {{3M} \over {4 \pi a_0^3}}{1 \over {\int_0^1 d\mu \, \xi_0(\mu)^3}}.
\end{equation}

For numerical convenience one may use the dimensionless density increment for the $i$-th spheroid:
 \begin{equation} 
\delta_i \equiv \delta \rho_i / \bar{\rho}
\end{equation}

As can be seen by examining Eqs. (10, 24-27), the dimensionless multipole moments can
be calculated using either the $\delta \rho_i$ or the $\delta_i$.  However, although the
moments are dimensionless, further scaling is necessary in order to achieve satisfactory numerical
accuracy.  Consider, for example, a model with $N=128$, having spheroids
with equally-spaced equatorial radii.  It then becomes necessary to consider
spheroids with $\lambda_i \sim 1/100$, so for example $J'_{100,20}$ has an integrand
$\sim 10^{-2 \times (-18)}$.  The resulting huge number is then multiplied by
$\sim  10^{-2 \times (+21)}$ in the corresponding term in Eq. (28).  To avoid
pointlessly multiplying and then dividing by large factors,
we rescale to new variables and parameters:
\begin{equation} 
\zeta_i(\mu) \equiv \xi_i(\mu) / \lambda_i ,
\end{equation}

\begin{equation} 
\tilde{J}_{i,2k} \equiv {J}_{i,2k} / \lambda_i^{2k} ,
\end{equation}

\begin{equation} 
\tilde{J}'_{i,2k} \equiv J'_{i,2k}  \lambda_i^{2k+1} ,
\end{equation}

Then
\begin{equation} 
\tilde{J}_{i,2k} = -\left({3 \over {2k+3}} \right)
\left( {{\delta_i \lambda_i^3 \int_0^1 d\mu \, P_{2k}(\mu) \, \zeta_i (\mu)^{2k+3} }
\over { \sum_{j=0}^{N-1}  \delta_j \lambda_j^3 \int_0^1 d\mu \, \zeta_j (\mu)^{3} } }\right)  ;
\end{equation}
for $k>1$
\begin{equation} 
\tilde{J}'_{i,2k} = -\left({3 \over {2-2k}} \right)
\left( {{\delta_i \lambda_i^3 \int_0^1 d\mu \, P_{2k}(\mu) \, \zeta_i (\mu)^{2-2k} }
\over { \sum_{j=0}^{N-1}  \delta_j \lambda_j^3 \int_0^1 d\mu \, \zeta_j (\mu)^{3} } }\right)  ,
\end{equation}
for $ k = 1$
\begin{equation} 
\tilde{J}'_{i,2} = -3
\left( {{\delta_i \lambda_i^3 \int_0^1 d\mu \, P_{2}(\mu) \ln[\zeta_i (\mu)] }
\over { \sum_{j=0}^{N-1}  \delta_j \lambda_j^3 \int_0^1 d\mu \, \zeta_j (\mu)^{3} } }\right)  ,
\end{equation}
and for $k = 0$
\begin{equation} 
\tilde{J}'_{i,0} = -{3 \over 2}
\left( {{\delta_i \lambda_i^3 \int_0^1 d\mu \, P_{2k}(\mu) \, \zeta_i (\mu)^{2} }
\over { \sum_{j=0}^{N-1}  \delta_j \lambda_j^3 \int_0^1 d\mu \, \zeta_j (\mu)^{3} } }\right)  ,
\end{equation}

We introduce dimensionless
planetary units of pressure ($P_{\rm pu}$), density
($\rho_{\rm pu}$) , and total potential ($U_{\rm pu}$), such that
\begin{equation} 
P \equiv {{GM^2} \over {a_0^4}} P_{\rm pu},
\end{equation}

\begin{equation} 
\rho \equiv {{M} \over {a_0^3}} \rho_{\rm pu},
\end{equation}

\begin{equation} 
U \equiv {{GM} \over {a_0}} U_{\rm pu}.
\end{equation}

Evaluating the total potential at the surface of the outermost Maclaurin
spheroid at the equator ($\mu=0$), we have

\begin{equation} 
U_{0, \rm pu} = 1 + {1 \over 2}q - \sum_{i=0}^{N-1}
\sum_{k=1}^{\infty} \tilde{J}_{i,2k} \lambda_i^{2k}  P_{2k}(0) .
\end{equation}

At the surface of each subsequent Maclaurin spheroid we have
\begin{align}
U_{j, \rm pu} &= - {1 \over \lambda_j} \left(
\sum_{i=j}^{N-1}
\sum_{k=0}^{\infty} \tilde{J}_{i,2k} ({\lambda_i / \lambda_j})^2k
  P_{2k}(0) \right. \nonumber\\
&+ \left.\sum_{i=0}^{j-1}
\sum_{k=0}^{\infty} \tilde{J}'_{i,2k} ({\lambda_j / \lambda_i})^{2k+1}
  P_{2k}(0) + \sum_{i=0}^{j-1} J''_{i,0} \lambda_j^3 \right)
+ {1 \over 2} q \lambda_j^2.
\end{align}
and at the center of the planet
\begin{equation} 
U_{\rm center, pu} = - \sum_{i=0}^{N-1}
\tilde{J}'_{i,2k} \lambda_i =  - \sum_{i=0}^{N-1} {J}'_{i,2k}.
\end{equation}

The shape $\zeta_0(\mu)$ of the surface of the planet is an equipotential
given by the solution to

\begin{equation} 
{1 \over \zeta_0} \left( 1 - \sum_{i=0}^{N-1}
\sum_{k=1}^{\infty} \tilde{J}_{i,2k} \lambda_i^{2k} \zeta_0^{-2k} 
  P_{2k}(\mu) \right)
+ {1 \over 3} q \zeta_0^2 [1 - P_2(\mu)]
= U_{0, \rm pu}.
\end{equation}

Correspondingly, the shape $\zeta_j(\mu)$ of the surface of the
$j$-th spheroid is an equipotential given by the solution to

\begin{gather} 
{ - {1 \over \zeta_j} \left(
\sum_{i=j}^{N-1}
\sum_{k=0}^{\infty} \tilde{J}_{i,2k} ({\lambda_i / \lambda_j})^{2k} \zeta_j^{-2k}
  P_{2k}(\mu)   
 + \sum_{i=0}^{j-1}
\sum_{k=0}^{\infty} \tilde{J}'_{i,2k} ({\lambda_j / \lambda_i})^{2k+1}
\zeta_j^{2k+1}  P_{2k}(\mu) \right. } \nonumber\\
{+ \left. \sum_{i=0}^{j-1} J''_{i,0} \lambda_j^3 \zeta_j^3 \right) 
+ {1 \over 3} q \lambda_j^3 \zeta_j^2 [1-P_2(\mu)]  
 + \left(
\sum_{i=j}^{N-1}
\sum_{k=0}^{\infty} \tilde{J}_{i,2k} ({\lambda_i / \lambda_j})^{2k}
  P_{2k}(0) \right. } \nonumber\\
{+ \left. \sum_{i=0}^{j-1}
\sum_{k=0}^{\infty} \tilde{J}'_{i,2k} ({\lambda_j / \lambda_i})^{2k+1}
  P_{2k}(0) + \sum_{i=0}^{j-1} J''_{i,0} \lambda_j^3 \right)
- {1 \over 2} q \lambda_j^3 =0 }.
\end{gather}

\subsection{Gaussian quadrature}

All of the foregoing expressions for the potential of $N$
concentric Maclaurin spheroids are
exact.  For practical applications, we are interested in finding the
potential as a multipole expansion to finite (say, thirtieth) degree,
corresponding to an upper limit at, say, $k_{\rm max} =15$.   For this purpose
one may numerically evaluate the angular integrals for the multipole moments using
$L > 2k_{\rm max}$ gaussian quadrature points.  For the examples presented
in this paper, we use $L=48$ gaussian quadrature points
$\mu_{\alpha}, \; \alpha= 1, 2, \mathellipsis L$ with corresponding
weights $w_{\alpha}, \; \alpha = 1, 2, \mathellipsis L$
 over the interval $0 < \mu < 1$.

Using initial guesses for the moments $\tilde{J}_{i,2k}$,  $\tilde{J}'_{i,2k}$, and  $J''_{i,0}$, one
solves Eqns. (50) and (51) for $\zeta_i(\mu_{\alpha})$ for $i=0, 1, \mathellipsis, N-1$ and
$\alpha = 1, 2, \mathellipsis, L$.  The solutions for these values are then used to evaluate
the gravitational moments by gaussian quadrature,
\begin{equation} 
\tilde{J}_{i,2k} \approx -\left({3 \over {2k+3}} \right)
\left( {{\delta_i \lambda_i^3 \sum_{\alpha=1}^L  w_{\alpha} \,P_{2k}(\mu_{\alpha}) \, \zeta_i (\mu_{\alpha})^{2k+3} }}
\over { \sum_{j=0}^{N-1}  \delta_j \lambda_j^3 \sum_{\alpha=1}^L  w_{\alpha}\, \zeta_j (\mu_{\alpha})^{3} } \right)  ,
\end{equation}
etc.

One then iterates between calculation of the level surface shapes
via Eqns. (50) and (51) and the gravitational moments
via Eqns. (40-43) until the difference between successive
iterations falls below a specified
tolerance.  For the purposes of achieving {\it Juno}-level
precision, about 30 such
iterations (over all $N$ spheroids) usually suffices.

\subsection{Calculation of the barotrope}

First, we calculate the density in each uniform layer; for the $j$-th layer
\begin{equation} 
\rho_{j, \, {\rm pu}} = {{\sum_{i=0}^j \delta_i} \over {\sum_{k=0}^{N-1} \delta_k \lambda_k^3
\sum_{\alpha=1}^L \zeta_k(\mu_{\alpha})^3}}.
\end{equation}

Next, we calculate the total potential $U_{\rm pu}$
on the outer surface, on each of the interfaces,
and at the center, using Eqs. (47-49). Since the density
is constant between interfaces, Eq. (31)
is trivially integrated to obtain the pressure at the bottom of the $j$-th layer:
\begin{equation} 
P_{j, \, {\rm pu}} = P_{j-1, \, {\rm pu}} + \rho_{j-1, \, {\rm pu}}
 (U_{j, \, {\rm pu}} - U_{j-1, \, {\rm pu}} ).
\end{equation}
Figure 4 shows an example of the resulting stair-step barotrope obtained
for a rotating Jupiter model with $N=32$ and a linear variation of density
with mean radius (the linear-density model is discussed further below).

\begin{figure}
\epsscale{.80}
\plotone{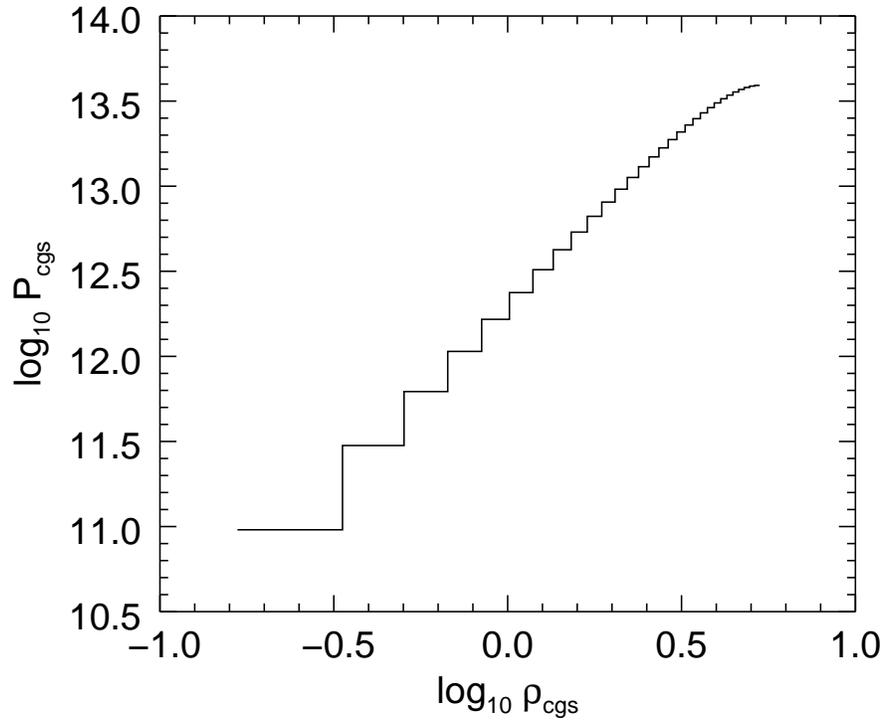}
\caption{Inferred variation of pressure vs. density (both in c.g.s. units)
in a CMS model of Jupiter with $N=32$, for an assumed linear variation of
mass density with mean radius.}
\label{fig4}
\end{figure}

\section{Comparison of CMS results with test cases}

\subsection{Linear density profile}

Results for a linear density model of Jupiter
using a fifth-order theory are tabulated in Table 3.1 of \citet{zt78}.
They adopt a mass-density profile which is linear in the
mean radius of a level surface rather than in its equatorial radius.
The mean radius $s_j$ of the $j$-th level surface relative to the
planetary mean radius is given by
\begin{equation} 
{{s_j} \over {s_0}} = \left({\int_0^1 d\mu \, \xi_j(\mu)^3} \over
 {\int_0^1 d\mu \, \xi_0(\mu)^3} \right)^{1/3} .
\end{equation}
If we arrange a constant increment $\delta_j$ in 
$\lambda_j$ (with constant $\Delta \lambda$),
we can make the resulting density linear
in $s/s_0$ by modifying the density increment for
each spheroid to
\begin{equation} 
\delta_{j, \, {\rm s}} = \delta_j {{\Delta s} \over {\Delta \lambda}}.
\end{equation}
The intervals $\Delta s$ must be computed iteratively.  Furthermore, 
\citet{zt78} expand their fifth-order theory in powers of the small
parameter 
\begin{equation} 
m = {{\omega^2 s_0^3} \over {GM}} ,
\end{equation}
with a fixed value of $m$ that differs slightly from the value obtained
from the value obtained for a more realistic Jupiter model.  Thus, the
CMS calculations must be also iterated to obtain a value for $m$ that matches
the one given by \citet{zt78}.

Table 1 presents a comparison of results for the linear-density model.
Agreement is excellent for $N=128$.  The inferred pressure-density
relation for $N=32$ was depicted in Fig. 4.

\subsection{Two-layer Maclaurin spheroids}

The relative simplicity and elegance of Maclaurin's theory for the single
spheroid disappears for $N=2$.  Nevertheless, one finds considerable
literature for the case $N=2$, dating back at least to \cite{dar03}.

First, it is useful to test the CMS theory by calculating the equipotential shape
of an interior interface in a Maclaurin spheroid of uniform density.  For
this test, I set $N=2$, $\lambda_1=0.5$, $\delta_0 = 1$, $\delta_1=0$.
I set $q$ equal to the Jovian value adopted in Paper I.  The converged CMS
model agrees exactly with results presented in Paper I, as it should.  Figure 5
shows the deviations of the outer and intermediate surfaces from ellipsoids of
revolution, with
$\delta \zeta = \zeta(\mu) - 1/ \sqrt{(1+ \ell^2 \mu^2)}$, where $\ell$
is related to $m$ by Maclaurin's result,
$m = {3 \over 2 \ell^3} [(3 + \ell^2) \arctan \ell - 3 \ell]$. 
Evidently the shape of the intermediate surface is
to high precision an ellipsoid of
revolution, homologous to the outer surface, as it is in
Maclaurin's analytic theory.

\begin{figure}
\epsscale{.80}
\plotone{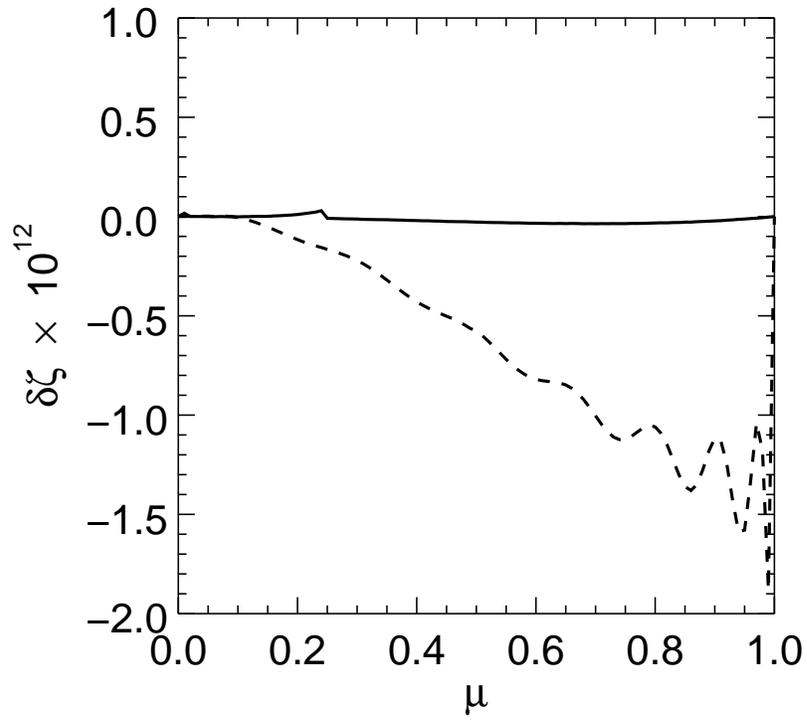}
\caption{Departure of outer surface (dashed) and intermediate surface (solid)
from an ellipsoid of revolution, for a classical Maclaurin spheroid.}
\label{fig5}
\end{figure}

Next, I compare my $N=2$ CMS results with the results of \citet{schu11}.
The Schubert et al. models are characterized by three parameters,
\begin{equation} 
Q_v = {{\int_0^1 d\mu \, \xi_1(\mu)^3} \over
 {\int_0^1 d\mu \, \xi_0(\mu)^3}}  ,
\end{equation}
the core-envelope density ratio $\rho_1/\rho_0$, and a dimensionless
rotation parameter
\begin{equation} 
\epsilon_2 = {\omega^2 \over
 {2 \pi G \rho_0} },
\end{equation}
all in my notation.  In the present paper I compare three models adopted by
\citet{schu11}: ``Mars'', ``Neptune'', and ``Uranus2''.  The quantities that
they compute for these models are $J_2$, and $E_0$ and $E_1$, respectively
the eccentricities of the outer surface and intermediate surface, defined by
\begin{equation} 
E = \sqrt{1 - (1 - e)^2} ,
\end{equation}
where the oblateness $e = 1 - \zeta(\mu=1)$.  The CMS calculations require
iteration to match the values of $Q_v$ and $\epsilon_2$.  Results are
presented in Tables 2-4.  While the values for ``Mars'' generally agree,
there are unexplained discrepancies for ``Neptune'' and ``Uranus2''.
The ``3rd order'' results of Schubert et al. agree with CMS results
to high precision, but their ``exact'' results differ by larger-than-expected
amounts.

\subsection{Polytrope of index one}

The polytrope of index one is defined by the barotrope
\begin{equation} 
P = K \rho^2 ,
\end{equation}
where the polytropic constant $K$ is chosen in the present application
to yield a model planet matched to Jupiter's mass and equatorial radius.  Rotating
planet models obeying this barotrope have been extensively studied \citep{zt78, WBH75}, so
it provides a rigorous test of the CMS method.

Moreover, the study presented in this section provides a useful
illustration of how, for
a chosen barotrope, one may choose CMS arrays of $\lambda_j$
and $\delta_j$ to yield a close match to the barotrope.

For a nonrotating $n=1$ polytrope, the density distribution is given by
\begin{equation} 
\rho = \rho_c \, {{\sin \pi \lambda} \over { \pi \lambda}} ,
\end{equation}
where $\rho_c$ is the central density.  To obtain a first approximation to
the $\delta$ distribution over the spheroids, we differentiate:
\begin{equation} 
{{d (\rho / \rho_c)} \over {d \lambda}} = {{\cos \pi \lambda} \over {\lambda}} -  {{\sin \pi \lambda} \over {\pi \lambda^2}} ,
\end{equation}
and we use this relation to obtain starting values of the $\delta_j$.

After obtaining a converged hydrostatic-equilibrium model for $N$ spheroids
with the above array of $\delta_j$,
one calculates the arrays $U_{j, \rm pu}$ and $P_{j, \rm pu}$.  Next one
calculates an array of desired densities $\rho_{j, \rm pu, \, desired}$
according to
\begin{equation} 
\rho_{j, \rm pu, \, desired} = \rho \left({1 \over 2}(P_{j+1} + P_j) \right) ,
\end{equation}
where $\rho (P)$ is the inverse of the adopted barotrope
$P(\rho )$.  Differencing the desired densities between layers then
gives a new array of $\delta_j$.  In general, it is necessary to scale the densities
so as to obtain the correct total mass of the CMS model.  This can be effected by
rewriting the barotrope as
\begin{equation} 
P = P(C \rho) ,
\end{equation}
where $C$ is a dimensionless factor.  For the polytrope of index one,
when one adopts a value of $C$ greater or less than one,
this is equivalent to redefining the value of $K$.

After obtaining a new converged CMS model, the process of adjusting the
densities to obtain a new array of  $\rho_{j, \rm pu, \, desired}$ ,
etc., continues until all changes in gravitational moments and
in the value of $C$ are
reduced to within a specified tolerance.  Because of the additional step of fitting the
barotrope, more iterations are required for convergence.  
Figure 6 shows the fitted and target $n=1$ barotrope of a converged 512-layer
CMS model of Jupiter. 

\begin{figure}
\epsscale{.80}
\plotone{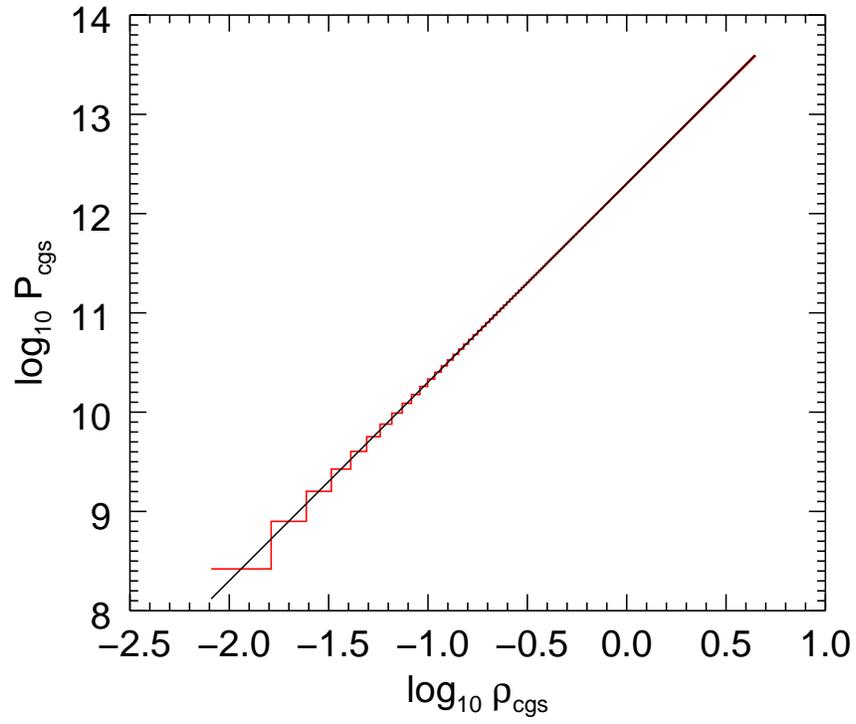}
\caption{Target (solid line) $n=1$ barotrope for a Jupiter model, and fitted $N=512$
CMS barotrope (stairstep).}
\label{fig6}
\end{figure}

The comparison models for the $n=1$ polytrope are (1) analytic expansions of
$J_2$, $J_4$, and $J_6$ to order $q^3$ \citep{zt78, WBH75},
and (2) a self-consistent-field calculation
of the rotating polytrope based on the analytic result that the interior
density can be expanded as a series of products of spherical harmonics and spherical
Bessel functions $j_n$ \citep{WBH75}.
 Table 6 presents results for $N=256$ and $N=512$ CMS
models along with the comparison models.  The results for $J_2$
for the  $N=512$ CMS model were still changing in the fifth figure after the
decimal point after fifteen iterations on the barotrope fit.

\subsection{Convergence considerations}

\citet{kzs13} have criticized the Maclaurin spheroid
approach employed in this paper, stating that the method of Paper I
is incomplete.  Further clarification is called for since the method
of Paper I is central to the CMS method.

Consider a Maclaurin spheroid of eccentricity $\ell$.  For its external potential
write, as usual,
\begin{equation} 
V_{\rm ext}(r,\mu)= {GM \over r}
\left[ 1 - \sum_{k=1}^{\infty} \left({a \over r} \right)^{2k} J_{2k} P_{2k}(\mu) \right]. 
\end{equation}
Where does this infinite-series expansion diverge for the Maclaurin spheroid? 
Evaluate it at the spheroid's pole, where $\mu = 1$ and $r^2 / a^2 = b^2 / a^2 =
1 / (1+\ell^2)$.  Substitute Eq. (10) of \citet{WBH12}.  We get

\begin{equation} 
V(b,1)= {GM \over b}
\left[ 1 - \sum_{k=1}^{\infty} 
{{3 (-1)^{1+k} \ell^{2k}} \over
{(2k+1) (2k+3)}} \right]. 
\end{equation}

 The
ratio of the $k+1$-th to the $k$-th term is

\begin{equation} 
{{t_{k+1}} \over {t_k}} = - \ell^2
{{(2k+1) (2k+3)} \over
{(2k+3) (2k+5)}}. 
\end{equation}

Therefore the series converges if $\ell^2 < 1$ or the oblateness $e < 1 - b/a = 1 - 
1/\sqrt{2} = 0.29289$, in agreement with the estimate of \citet{kzs13}.
The corresponding $m = m_{\rm crit} = 0.212389$ and $q = q_{\rm crit} = 0.424778$.
These values are far larger than the parameters of any known planet.  Note, by the
way, that the point of bifurcation for the Maclaurin-Jacobi ellipsoid sequence is
at a somewhat larger $\ell_{\rm bifurc} = 1.39$, corresponding to $m_{\rm bifurc} = 0.280$
and $q_{\rm bifurc} = 0.669$.

Paper I shows that for a Maclaurin spheroid with Jupiter's $q = 0.089$,
my method gives for the shape of the spheroid's surface a numerical
result that differs no more than
a few parts in $10^{13}$ from the exact Maclaurin shape.  Note that
Figure 5 of this paper shows similarly-small departures at the outer
surface and on an intermediate surface.
Thus, for this value of $q$, any neglected terms in Equation (66)
will not exceed $\sim 10^{-12}$ of the included terms.

Repeating the calculation for a Maclaurin spheroid with a Saturn-like $q = 0.155$, I
obtain results shown in Figure 7 for the relative difference of the spheroid's surface
radius from the exact Maclaurin ellipsoid shape.  The departures are,
in absolute terms, no more
than a few centimeters, and would have no significance whatsoever for practical
models of Saturn's gravity field.  Moreover, the real Saturn is much less oblate than the
Maclaurin model, so the departures of a CMS model from an ``exact'' model will be smaller still.
 
As in Paper I, the general CMS method relies upon the requirement that
the external multipole expansion (66) of a given spheroid's potential
converges at all points on the spheroid's surface.  First, on
a sphere of radius $r = a_0$, the expansion converges.
To see this, note that $J_{2k} \, \sim \, (-1)^{k+1} q^k$ for a
uniformly-rotating body in hydrostatic equilibrium.
Thus, on the sphere $\xi_0=1$, the ratio of the $k+1$-th term
to the $k$-th term is $\sim \, -q$, so for $q \, < \, 1$ the series
converges.

Next we examine the series convergence at the pole, $\mu \, = \, 1$,
where $\xi_0 \, = \, 1-e$, where $e \, \sim \, q$ is the oblateness.
At $\mu \, = \, 1$, the ratio of the $k+1$-th term
to the $k$-th term is $\sim \, -q/(1-e)^2$.  Thus, as
long as $q \, < \, C' (1-e)^2$ (where $C'$ is a constant
of order unity whose precise value depends on the
barotrope) the series converges. As discussed, the $q$ values for Jupiter and
Saturn are such that the convergence criterion
is well satisfied (as numerically demonstrated for
the test cases).  See also a relevant discussion in Section 38 of
\citet{zt78}.

\begin{figure}
\epsscale{.80}
\plotone{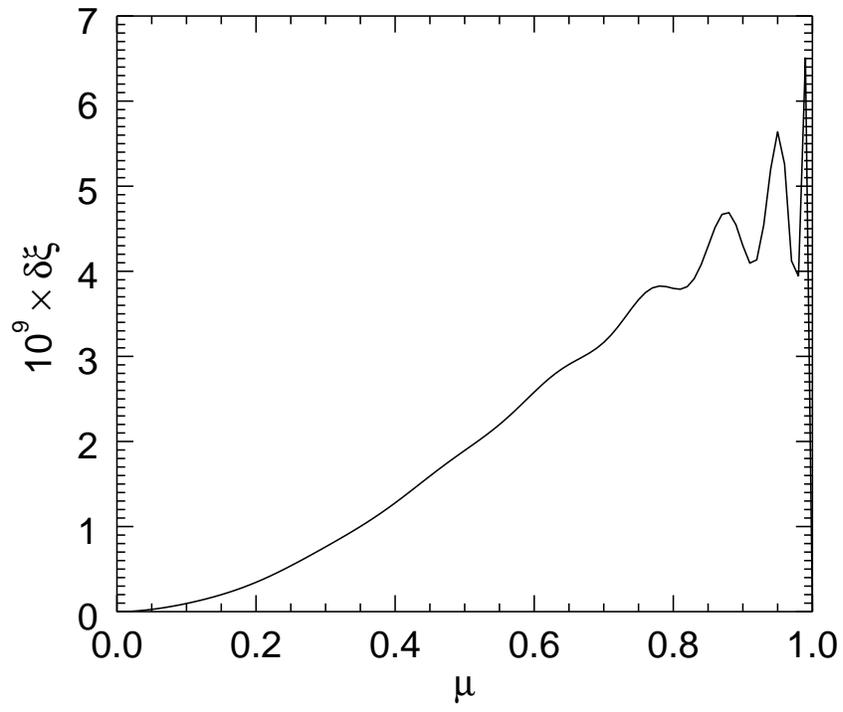}
\caption{Calculation of the difference $\delta \xi = \xi_{\rm CMS} - \xi_{\rm exact}$
for a Maclaurin spheroid with Saturn's mass, mean density, and rotation rate.}
\label{fig7}
\end{figure}

\section{Practical application to analysis of gravity data}

The CMS analysis technique presented here can be vectorized for efficiency,
although no significant effort has been made to do so at this point.  For
practical computations it will probably be necessary to further increase $N$
and to increase the number of iterations on the barotrope fit, in order
to match the theoretical results to the expected precision of spacecraft
measurements.

Further iteration loops will be required if a subset of the calculated $J_{2k}$
are to be fitted to observed values.  Adjustable parameters might include:
(1) the mass and density of a discrete core at the planet's center,
(2) chemical and density discontinuities at various layers, and
(3) modifications to the assumed barotrope (crudely illustrated in this
paper with the scale factor $C$).

As is obvious, there exists an infinity of possible arrangements of
spheroids which can be fitted to a finite set of gravity data.  Thus, a unique
inversion cannot be achieved.  However, application of specific
physically-based barotropes and cosmochemical considerations
can lead to the most realistic interior models.

\section{Conclusion}

One can further generalize the CMS method in two directions.  First, 
in addition to the rotational potential $Q$ one may introduce a tidal
potential from a satellite.  The resulting tidal perturbing potential $Q_{\rm tid}$
will be a function of two angular variables, $\mu$ and $\phi$, where $\phi$
is the angle from the sub-satellite longitude.  Since $Q_{\rm tid}$ will excite
both zonal and tesseral gravity harmonics, evaluation of the response
on all CMS surfaces will require two-dimensional rather than one-dimensional
integrals.  However, there appears to be no practical barrier to evaluating
such integrals using two-dimensional
gaussian quadrature (to be sure, at the cost of more
computing time).

Second, one can investigate the effects of differential rotation
on cylinders (for related investigations, see \citet{kong12} and \citet{WBH82}).

\acknowledgments

This research is supported by the {\it Juno} project under Subcontract 699053X
from the Southwest Research Institute to the University of Arizona.

\clearpage


\begin{deluxetable}{rrrrrrrr}
\tablecolumns{3}
\tablewidth{0pc}
\tablecaption{Linear Density Model}
\tablehead{
\colhead{Quantity} & \colhead{ZT\tablenotemark{a} $5^{\rm th}-$order theory}   & \colhead{CMS theory ($N=128$)}  }
\startdata
$q$\phn  & \nodata \phn \phn &  $0.088822426$ \phn \\
$m$\phn  & $0.0830$\phn \phn &  $0.082999915$ \phn \\
$J_2 \times 10^2$\phn  & $1.4798$\phn \phn & $1.4798138$  \phn \\
$-J_4 \times 10^4$\phn  & $5.929$\phn  \phn & $5.9269129$ \phn   \\ 
$J_6 \times 10^5$\phn  & $3.497$\phn  \phn &  $3.4935680$ \phn \\ 
$-J_8 \times 10^6$\phn  & $2.52$\phn  \phn &  $2.5493209$ \phn \\ 
$J_{10} \times 10^7$\phn  & $2.4$ \phn  &  $2.1308951$ \phn \\ 
$-J_{12} \times 10^8$\phn  & \nodata \phn  &  $1.9564143$ \phn \\ 
$J_{14} \times 10^9$\phn  & \nodata \phn  &  $1.9237724$ \phn \\ 
\enddata
\tablenotetext{a}{\citet{zt78} }
\end{deluxetable}


\begin{deluxetable}{rrrrrrrr}
\tablecolumns{5}
\tablewidth{0pc}
\tablecaption{``Mars'' {\bf Note.} ``3rd order'' and ``exact'' values from Schubert et al. (2013)}
\tablehead{
\colhead{Quantity} & \colhead{value}  & \colhead{``3rd order''}   & \colhead{``exact''} & \colhead{CMS ($N=2$)}  }
\startdata
$Q_v$\phn  & 0.125 \phn \phn &   \phn &   \phn &   \phn \\
$\rho_0 / \rho_1 $\phn  & 0.486 \phn \phn &   \phn &   \phn &   \phn \\
$\epsilon_2$\phn  & 0.00347 \phn \phn &   \phn &   \phn &   \phn \\
$q$\phn  & 0.0046205430 \phn \phn &   \phn &   \phn &   \phn \\
$J_2 \times 10^6$\phn  & \nodata \phn & $1823.18$\phn \phn &   1823.1 \phn & 1823.1832  \phn \\
$E_1$ \phn  & \nodata \phn & $0.0888747$\phn \phn &   $0.088859$ \phn & $0.088874693$  \phn \\
$E_0$ \phn  & \nodata \phn & $0.100295$\phn \phn &   $0.10030$ \phn & $0.10029471$  \phn \\
\enddata
\end{deluxetable}


\begin{deluxetable}{rrrrrrrr}
\tablecolumns{5}
\tablewidth{0pc}
\tablecaption{``Neptune'' {\bf Note.} ``3rd order'' and ``exact'' values from Schubert et al. (2013)}
\tablehead{
\colhead{Quantity} & \colhead{value}  & \colhead{``3rd order''}   & \colhead{``exact''} & \colhead{CMS ($N=2$)}  }
\startdata
$Q_v$\phn  & 0.091125 \phn \phn &   \phn &   \phn &   \phn \\
$\rho_0 / \rho_1 $\phn  & 0.157334 \phn \phn &   \phn &   \phn &   \phn \\
$\epsilon_2$\phn  & 0.0254179 \phn \phn &   \phn &   \phn &   \phn \\
$q$\phn  & 0.026207112 \phn \phn &   \phn &   \phn &   \phn \\
$J_2 \times 10^6$\phn  & \nodata \phn & $6188.92$\phn \phn &   $6241.0$ \phn & 6188.9267  \phn \\
$E_1$ \phn  & \nodata \phn & $0.143515$\phn \phn &   $0.15147$ \phn & $0.14351534$  \phn \\
$E_0$ \phn  & \nodata \phn & $0.209658$\phn \phn &   $0.21019$ \phn & $0.20965898$  \phn \\
\enddata
\end{deluxetable}


\begin{deluxetable}{rrrrrrrr}
\tablecolumns{5}
\tablewidth{0pc}
\tablecaption{``Uranus2'' {\bf Note.} ``3rd order'' and ``exact'' values from Schubert et al. (2013)}
\tablehead{
\colhead{Quantity} & \colhead{value}  & \colhead{``3rd order''}   & \colhead{``exact''} & \colhead{CMS ($N=2$)}  }
\startdata
$Q_v$\phn  & 0.0563272 \phn \phn &   \phn &   \phn &   \phn \\
$\rho_0 / \rho_1 $\phn  & 0.0791231 \phn \phn &   \phn &   \phn &   \phn \\
$\epsilon_2$\phn  & 0.0318902 \phn \phn &   \phn &   \phn &   \phn \\
$q$\phn  & 0.029581022 \phn \phn &   \phn &   \phn &   \phn \\
$J_2 \times 10^6$\phn  & \nodata \phn & $5680.32$\phn \phn &   $5801.4$ \phn & $5680.3242$  \phn \\
$E_1$ \phn  & \nodata \phn & $0.115655$\phn \phn &   $0.14160$ \phn & $0.11565564$  \phn \\
$E_0$ \phn  & \nodata \phn & $0.213648$\phn \phn &   $0.21473$ \phn & $0.21364898$  \phn \\
\enddata
\end{deluxetable}


\begin{deluxetable}{rrrrrrrr}
\tablecolumns{5}
\tablewidth{0pc}
\tablecaption{Polytrope $n=1$}
\tablehead{
\colhead{Quantity} & \colhead{3rd order theory}   & \colhead{ $j_n$ expansion} & \colhead{CMS ($N=256$)}  & \colhead{CMS ($N=512$)}}
\startdata
$q$\phn \phn \phn  & $0.089195487$ \phn  &  $0.089195487$ \phn & $0.089195487$  \phn & $0.089195487$  \phn  \\
$J_2 \times 10^2$\phn  & $1.3994099$\phn \phn& $1.3988511$ \phn   & $1.3991574$  \phn & $1.3989253$  \phn \\
$-J_4 \times 10^4$\phn  & $5.3871087$\phn  \phn & $5.3182810$ \phn  & $5.3203374$ \phn & $5.3187997$ \phn    \\ 
$J_6 \times 10^5$\phn  & $3.9972442$\phn  \phn & $3.0118323$ \phn  &  $3.0133819$ \phn &  $3.0122356$ \phn \\ 
$-J_8 \times 10^6$\phn  & \nodata \phn  \phn & $2.1321157$ \phn  &  $2.1334136$ \phn &  $2.1324628$ \phn \\ 
$J_{10} \times 10^7$\phn  & \nodata  \phn  & $1.7406710$ \phn  &  $1.7418428$ \phn &  $ 1.7409925$ \phn \\ 
$-J_{12} \times 10^8$\phn  & \nodata \phn  & $1.5682179$ \phn  &  $1.5693324$  \phn &  $1.5685327$ \phn \\ 
$J_{14} \times 10^9$\phn  & \nodata \phn  & $1.5180877$ \phn  &  $1.5191923$  \phn &  $1.5184156$ \phn \\ 
\enddata

\end{deluxetable}

\end{document}